\documentclass[conference]{IEEEtran}
\IEEEoverridecommandlockouts
\usepackage{cite}
\usepackage{amsmath,amssymb,amsfonts,latexsym}
\usepackage{braket}
\usepackage{algorithmic}
\usepackage{graphicx}
\usepackage{textcomp}
\usepackage{xcolor}
\usepackage{euscript,mathrsfs}
\usepackage{mathtools}
\usepackage{yhmath}
\usepackage{url}
\usepackage{empheq}
\usepackage{booktabs}
\usepackage{hyperref}
\usepackage{lipsum}
\usepackage[all]{xy}
\usepackage{tikz-cd}
\usepackage{cuted}
\usepackage{tikz}
\usetikzlibrary{intersections,shapes.arrows}

\usepgfmodule{nonlineartransformations}
\def\fluttertransform{%
    \pgfgetlastxy\x\y
    \pgfpoint{\x+sin(\y)}{\y+sin(\x)*(30-\x/2)+\x/10}
}

\def \vt{\vartheta}

\def \d{\delta}

\def \s{\sigma}
\def \p{\partial}

\def\wtilde{\widetilde}
\def\mcal{\mathcal}

\def \Ud{U^{\dagger}}
\def \Vd{V^{\dagger}}

\DeclareMathOperator*{\argmin}{arg\,min}

\def\BibTeX{{\rm B\kern-.05em{\sc i\kern-.025em b}\kern-.08em
    T\kern-.1667em\lower.7ex\hbox{E}\kern-.125emX}}

\makeatletter
\newcommand{\linebreakand}{%
  \end{@IEEEauthorhalign}
  \hfill\mbox{}\par
  \mbox{}\hfill\begin{@IEEEauthorhalign}
}
\makeatother

\begin{document}

\title{Transfer Learning Analysis of Variational Quantum Circuits \thanks{This work is supported by Laboratory Directed Research and Development Program \#24-061 of Brookhaven National Laboratory and National Quantum Information Science Research Centers, Co-design Center for Quantum Advantage (C2QA) under Contract No. DE-SC0012704. The views expressed in this article are those of the authors and do not represent the views of Wells Fargo. This article is for informational purposes only. Nothing contained in this article should be construed as investment advice. Wells Fargo makes no express or implied warranties and expressly disclaims all legal, tax, and accounting implications related to this article.\\ The code of this paper can be found at \url{https://github.com/QC-AI/VQC_Transfer_Learning.git}.}}

\author{
\IEEEauthorblockN{Huan-Hsin~Tseng}
\IEEEauthorblockA{\textit{AI \& ML Department} \\
\textit{Brookhaven National Laboratory}\\
Upton NY, USA  \\
htseng@bnl.gov}
\and
\IEEEauthorblockN{Hsin-Yi~Lin}
\IEEEauthorblockA{\textit{Department of Mathematics} \\
\textit{and Computer Science} \\
\textit{Seton Hall University}\\
South Orange NJ, USA \\
hsinyi.lin@shu.edu}
\and
\IEEEauthorblockN{Samuel Yen-Chi~Chen}
\IEEEauthorblockA{\textit{Wells Fargo} \\
New York NY, USA \\
yen-chi.chen@wellsfargo.com}
\and 
\IEEEauthorblockN{Shinjae~Yoo}
\IEEEauthorblockA{\textit{AI \& ML Department} \\
\textit{Brookhaven National Laboratory}\\
Upton NY, USA \\
syjoo@bnl.gov}
}

\maketitle

\begin{abstract}
This work analyzes transfer learning of the Variational Quantum Circuit (VQC). Our framework begins with a pretrained VQC configured in one domain and calculates the transition of 1-parameter unitary subgroups required for a new domain. A formalism is established to investigate the adaptability and capability of a VQC under the analysis of loss bounds. Our theory observes knowledge transfer in VQCs and provides a heuristic interpretation for the mechanism. An analytical fine-tuning method is derived to attain the optimal transition for adaptations of similar domains.
\end{abstract}


\section{Introduction}
Quantum Computing (QC) has demonstrated the ability to address complex computational challenges that are intractable for classical computers. Leveraging unique quantum properties, such as superposition and entanglement, quantum computers exhibit significant advantages over classical systems in solving certain tasks~\cite{nielsen2010quantum}. Although current quantum devices are constrained by imperfections and noise at current stage, a hybrid quantum-classical computing framework has been proposed to harness potential quantum advantages~\cite{abbas2021power, caro2022generalization, du2020expressive}. Variational Quantum Algorithms (VQAs)~\cite{bharti2022noisy} represent a class of hybrid algorithms in which quantum parameters are optimized with classical methods, such as gradient descent. VQAs enable architectures like Quantum Neural Networks (QNNs) to perform a variety of Machine Learning (ML) tasks, including classification~\cite{chen2021end, qi2023qtnvqc, mitarai2018quantum, chen2022quantumCNN,lin2024quantumGRADCAM}, time-series prediction~\cite{chen2022quantumLSTM,chen2024QFWP,chehimi2024FedQLSTM}, natural language processing~\cite{li2023pqlm, yang2022bert, di2022dawn, stein2023applying}, reinforcement learning~\cite{chen2020VQDQN,chen2022variationalQRL,chen2023QLSTM_RL,skolik2022quantum,jerbi2021parametrized,yun2022quantum} and neural network model compression \cite{liu2024training,liu2024qtrl,liu2024federated_QT,lin2024_QT_DeepFake,lin2024QT_LSTM}.

Driven by different computing mechanisms, QC model interpretability~\cite{lin2024quantum} and transparency are intriguing topics to be investigated for understanding how quantum systems process information and contribute to optimizations. This direction has attracted much attention lately, primarily due to extensive ML applications and the continuing development of policy and regulation. With the rapid growth of QC, unveiling deeper insights into quantum operations will be essential.

On the other hand, Transfer learning~\cite{pan2009survey, he2016deep, houlsby2019parameter} is a well-established technique in classical ML that uses prior knowledge to enhance model performance and reduces resource requirements. Naturally, this empirical approach is introduced in QML to offer similar advantages. An earlier study~\cite{mari2020transfer} explored transfer learning in hybrid classical-quantum neural networks, proposing VQCs as fine-tuning layers to replace classical ones in the pretrained network. A major distinction with our approach is that \cite{mari2020transfer} assumes a \emph{purely classical} pretrained model where classical layers are later swapped with VQCs for fine-tuning, while we consider fine-tuning resumed from a pretrained VQC. Consequently, in \cite{mari2020transfer}, quantum gate parameters are initialized from scratch, bearing no memory in regard to previous data. In contrast, we leverage the optimal quantum parameters from pretraining, preserving domain knowledge to digest new data.

It is noticed that \cite{tseng2023interpretations} investigates the transfer learning mechanisms of classical neural networks, illustrating how prior knowledge is transferred to new data domains. However, to analyze VQC, the required techniques and approaches are fundamentally different, which indicates the distinct behaviors and unique characteristics inherent to QML.

Our contribution is twofold: (1) We derive an analytical solution for adjusting VQC model parameters between similar datasets, which yields global optimal under specific data conditions. This approach is particularly efficient compared to the conventional gradient descent or the parameter shift rule~\cite{mari2020transfer}. (2) Our theoretical analysis reveals the underlying transfer learning mechanism of VQCs, providing insights into its physical implications.

\section{Background}

\subsection{Variational Quantum Circuits}

The Variational Quantum Circuit (VQC) has the design to fit and learn from external signals with tunable parameters in quantum gates. The construction of 1-qubit VQC is succinctly introduced in \cite{mitarai2018quantum}. VQCs of multiple qubits can be extended via tensor products with the building block below. 

Let the 1-qubit Hilbert space be $\mcal{H} = \mathbb{C}^2$ with a basis $\ket{0}, \ket{1}$ such that a spin-$\frac{1}{2}$ particle wave function $\ket{\psi} \in \mcal{H}$ can be decomposed as $\ket{\psi} = a \ket{0} + b \ket{1}$ for some $a, b \in \mathbb{C}$.

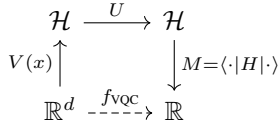
\begin{figure}[th]
\begin{center}
\begin{tikzcd}
\mcal{H} \arrow[r, "U"] & \mcal{H} \arrow[d, "M = \langle \cdot | H | \cdot \rangle"]\\
\mathbb{R}^d \arrow[u, "V(x)"] \arrow[r, dashed, "f_{\text{VQC}}"]
& \mathbb{R}
\end{tikzcd}
\end{center}
\caption{The diagram of a VQC with a classical input $x \in \mathbb{R}^d$ received.}\label{fig: VQC}
\end{figure}

Given classical data $\mcal{D} = \{ \left( x^{(i)}, y^{(i)} \right) \}_{i=1}^N$ with input $x^{(i)} \in \mathbb{R}^d$ and ground truth $y^{(i)} \in \mathbb{R}$ of sample index $i$, where the index may sometimes be neglected for simplicity when a context is clear. Denote all unitary transformations of $\mcal{H}$ by $\mcal{U}(\mcal{H})$. A VQC is a function $f: \mathbb{R}^d \to \mathbb{R}$ of three components $(V, U, M)$, $f = M \circ U \circ V$, see Fig.~\ref{fig: VQC}, where: $V : \mathbb{R}^d \to \mcal{U}(\mcal{H})$ is called an \emph{encoding} function converting a classical input $x$ into a quantum state $x \mapsto V(x) \ket{\psi_0}$ from an initial $\ket{\psi_0} \in \mcal{H}$. The encoding unitary $V(x)$ is typically a choice of circuit sequence $\{V_1, \ldots, V_k \}$ such that $ V := V_k \circ \cdots \circ V_j \circ \cdots \circ V_1$ with each $V_j: \mathbb{R}^d \to \mcal{U}(\mcal{H})$.

The \emph{variational} function $U \in \mcal{U}(\mcal{H})$ is the main modeling engine of VQC formed by a sequence of gates $\{ U_1, \ldots, U_L \}$ with tunable parameters $\theta = \{ \theta_1, \ldots, \theta_L\}$ such that $U(\theta) = U_L (\theta_L) \cdots \circ U_{\ell}(\theta_{\ell}) \circ \cdots  U_1(\theta_1)$. The conventional choice of $\phi \mapsto U_{\ell}(\phi)$ is usually a 1-parameter Lie subgroup in $\mcal{U}(\mcal{H})$.

Finally, a \emph{measurement} $M: \mathcal{H} \to \mathbb{R}$ is a function to collapse a quantum state by the choice of a Hermitian operator $H: \mathcal{H} \to \mathcal{H}$ such that $M(\psi) := \bra{\psi} H \ket{\psi}$. The property that $M(\psi) \in \mathbb{R}$ is due to $H^{\dagger} = H$ to mimic the output of a classical neural network. Together, a VQC has the form,
\begin{equation}\label{E: VQC output}
x \mapsto f (x) = \bra{\psi_0} \Vd(x) \Ud(\theta) \, H \, U(\theta) V(x) \ket{\psi_0}
\end{equation}
where $\Vd$ is the adjoint operator of $V \in \mcal{U}(\mcal{H})$. The exact choice of the $U$, $V$ for this study is specified in Sec.~\ref{subsec: VQC fine-tune error}.

\subsection{An adjoint action approximation}\label{subsec: Adjoint approx}
Recall that the adjoint action defined on a Lie group $G$: $Ad_g: G \to G$ by $Ad_g(h) = g h g^{-1}$ leads to an associated group homomorphism $\mathfrak{ad}: G \to GL(\mathfrak{g})$ by $g \mapsto (Ad_g)_{*, e}$, where $\mathfrak{g}$ is the Lie algebra of $G$ and $(Ad_g)_{*, e}: \mathfrak{g} \to \mathfrak{g} $ denotes the differential of $Ad_g$ at the identity $e \in G$. We know that for $A, B \in \mathfrak{g}$, the differential $\mathfrak{ad}_{*,e}: \mathfrak{g} \to \mathfrak{gl}(\mathfrak{g})$ at $e$ gives,
\begin{equation}\label{E:ad Lie}
(\mathfrak{ad}_{*,e})(A)(B) = \frac{\p^2}{\p s \p t} \left( e^{tA} \, e^{sB} \, e^{-tA} \right)\Bigr|_{s,t=0} = [A, B]
\end{equation}
where $[\cdot, \cdot]$ is the Lie bracket of $\mathfrak{g}$. Under a matrix group $G$, (\ref{E:ad Lie}) is reduced to
\[
\frac{d}{dt} \left( e^{tA} B e^{-tA} \right)\Bigr|_{t=0} = [A, B] = AB - BA
\]
which tells us,
\begin{equation}\label{E:ad approx}
 e^{tA} B e^{-tA} = B + t [A, B] + \mathscr{O}(t^2)
\end{equation}
In the event of $G = \mcal{U}(\mcal{H})$, we can take $A = i \s$ with $\s \in \mcal{P}$ where $\mcal{P}:= \{ \s_1, \s_2, \s_3 \}$ is the collection of Pauli matrices. Then (\ref{E:ad approx}) gives,
\begin{equation}\label{E:ad approx 2}
 e^{it \s} B e^{-i t\s} = B + it [\s, B] + \mathscr{O}(t^2)
\end{equation}
This can be viewed as an adjoint action approximation, which allows us to analyze VQC circuit transitions under transfer learning.

\section{Quantum Transfer Learning}

\subsection{VQC Learning in Two Domains}
A VQC learns to fit one dataset $\mathcal{D}$ by finding a proper parameter set $\theta$ minimizing an objective function,
\begin{equation}\label{E: source loss}
L(\theta; \mcal{D}) = \sum_{i=1}^N \| \braket{H} (x^{(i)}; \theta) - y^{(i)} \| ^2
\end{equation}
where $\braket{H} (x^{(i)}; \theta)$ is also used to denote $f(x)$ in (\ref{E: VQC output}) to emphasize the dependency on $H, x^{(i)}$, and $\theta$ under a fixed $\ket{\psi_0}$. Note that the \emph{optimal parameters} in script $\vt := \argmin_{\theta} L(\theta; \mcal{D})$ is to be distinguished from an arbitrary parameter set $\theta$.

\begin{figure}[htbp]
    \centering 
\begin{tikzpicture}[scale=1]
\begin{scope}[yshift=20mm]
\pgftransformnonlinear{\fluttertransform}
\draw [fill=blue!20] plot [smooth cycle]
coordinates {(-1.14,-1)(-0.84, -.18) (-0.04, 0.3) (2.24, 0) %
(4.48, -0.56) (4.48, -1.46) (3.38,-1.84)(0.38, -1.28)};
\end{scope}

\coordinate (p1) at (0.1, 1.6);
\coordinate (p2) at (1.4, 1.9);
\coordinate (p3) at (2.5, 1.05);
\coordinate (p4) at (3.8, 1.1);

\filldraw (5., 1.5) node[left] {$\mathcal{U}(\mathcal{H})$};

\filldraw (p1) circle (1pt) node[above] {\footnotesize{$U(\vartheta)$}};
\filldraw (p2) circle (1pt) node[above] {\footnotesize{$U(\theta^{(1)})$}};
\filldraw (p3) circle (1pt) node[below] {\footnotesize{$U(\theta^{(2)})$}};
\filldraw (p4) circle (1pt) node[below] {\footnotesize{$U(\widetilde{\vartheta})$}};

\draw[-, dashed] (p1) -- (p2) -- (p3) -- (p4);
\draw[-, red] (p1) -- (3.75, 1.1);
\end{tikzpicture}
\caption{A VQC finetuning from pretrained $\vt \to \wtilde{\vt}$ for domain adaptation. The dashed lines denote conventional search over new training iterations. The red line represents our 1-step analytical solution introduced in Sec.~\ref{subsec: VQC fine-tune error}.}
\label{fig: Unitary param transition}
\end{figure}
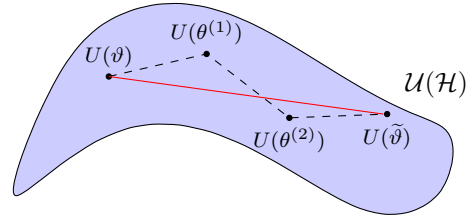

Given another domain data $\wtilde{\mcal{D}} = \{(\wtilde{x}^{(i)}, \wtilde{y}^{(i)}) \}_{i=1}^N$ of inputs $\wtilde{x}^{(i)} \in \mathbb{R}^d$ and labels $\wtilde{y}^{(i)} \in \mathbb{R}$ similar to a previous domain (data) $\mathcal{D}$, we consider the \emph{quantum transfer learning} by seeking new optimal $\wtilde{\vt}$ based on previous optimal $\vt$ learned in $\mathcal{D}$, see Fig.~\ref{fig: Unitary param transition}. Such a process is also called \emph{finetune} or \emph{domain adaptation}. In the context of ML, $\mcal{D}$ is also called the \emph{source domain} and $\wtilde{\mcal{D}}$ the \emph{target domain} with $\vt$ called the \emph{pretrain (model) parameters}. 

\subsection{VQC Finetune Error Estimates}\label{subsec: VQC fine-tune error}

In general, it is difficult to analyze the required parameter change $\vt \to \wtilde{\vt}$. However, to derive a glimpse of VQC transfer ability, we utilize several techniques for the estimation. First, we observe the target domain loss,
\begin{equation}\label{E: target loss}
L(\wtilde{\vt}; \wtilde{\mcal{D}}) = \sum_{i=1}^N \| \braket{H} (\wtilde{x}^{(i)}; \wtilde{\vt}) - \wtilde{y}^{(i)} \| ^2
\end{equation}
with
\begin{equation}
\braket{H} (\wtilde{x}^{(i)}; \wtilde{\vt}) := \bra{\psi_0} \Vd(\wtilde{x}^{(i)}) \Ud(\wtilde{\vt}) \, H \, U(\wtilde{\vt}) V(\wtilde{x}^{(i)}) \ket{\psi_0}
\end{equation}
For the convenience of analysis, we assume two domain samples of $\mcal{D}$ and $\wtilde{\mcal{D}}$ are rearranged via domain alignment~\cite{courty2016optimal, lin2021unsupervised} such that $\| (\wtilde{x}_i, \wtilde{y}_i) - (x_i, y_i) \| \leq \| (\wtilde{x}_i, \wtilde{y}_i) - (x_j, y_j) \|$ for all $j \neq i$. 
Then we may define $\d x^{(i)} := \wtilde{x}^{(i)} - x^{(i)}$, $\d y^{(i)} = \wtilde{y}^{(i)} - y^{(i)} $ and expand (\ref{E: target loss}) as,
\begin{multline}\label{E:target loss 2}
L(\wtilde{\vt}; \wtilde{\mcal{D}}) = \sum_{i=1}^N \Big\| \Big( \braket{H} (\wtilde{x}^{(i)}; \wtilde{\vt}) - \braket{H} (x^{(i)}; \vt) \Big) - \d y^{(i)} +\\
\big(\braket{H} (x^{(i)}; \vt) - y^{(i)} \big) \Big\|^2
\end{multline}
Assume the VQC architecture to learn the two domains is fixed as the following form,
\begin{equation}\label{E: VQC architecture}
\begin{aligned}
V(x) := V_d(x_d) \cdots V_1(x_1), \quad \left( V_j(x_j) = e^{-\frac{i}{2} x_j \s_{k_j}} \right)\\
U(\vt) := U_L(\vt_L) \cdots U_1(\vt_1), \quad \left( U_{\ell}(\vt_{\ell}) = e^{-\frac{i}{2} \vt_{\ell} \s_{k_{\ell}}} \right)
\end{aligned}
\end{equation}
with indices $k_j, k_{\ell} \in \{1, 2, 3\}$, $j=1, \ldots, d$ and $\ell = 1, \ldots, L$, $\vt = (\vt_1, \ldots, \vt_L)$, and $x = (x_1, \ldots, x_d)$ denoting a $d$-dim input with sample index ignored. Note the multiplication order in (\ref{E: VQC architecture}) has the \emph{earliest} action starting from the \emph{right}. By (\ref{E: VQC architecture}), we also have,
\begin{equation}\label{E: circuit connection}
\begin{aligned}
    V_j(\wtilde{x}_j) &= e^{-\frac{i}{2} \d x_j \s_{k_j}} \cdot V_j(x_j) =  V_j(x_j) \cdot e^{-\frac{i}{2} \d x_j \s_{k_j}}\\
    U_{\ell}(\wtilde{\vt}_{\ell}) &= e^{-\frac{i}{2} \d \vt_{\ell} \s_{k_{\ell}}} \cdot U_{\ell}(\vt_{\ell}) = U_{\ell}(\vt_{\ell}) \cdot e^{-\frac{i}{2} \d \vt_{\ell} \s_{k_{\ell}}}
\end{aligned}
\end{equation}
The last equalities in (\ref{E: circuit connection}) stress that the commutativity exists. We expand the first two terms grouped in (\ref{E:target loss 2}) as follows,
\begin{multline}\label{E: target loss 3}
 \braket{H} (\wtilde{x}; \wtilde{\vt}) - \braket{H} (x; \vt) =\\
    \Bigl< \mathfrak{ad}_{\Vd(\wtilde{x})} \mathfrak{ad}_{\Ud(\wtilde{\vt})} (H) \Bigr>  - \Bigl< \mathfrak{ad}_{\Vd(x)} \mathfrak{ad}_{\Ud(\wtilde{\vt})} (H) \Bigr> \\ 
    + \Bigl< \mathfrak{ad}_{\Vd(x)} \mathfrak{ad}_{\Ud(\wtilde{\vt})} (H) \Bigr>  - \Bigl< \mathfrak{ad}_{\Vd(x)} \mathfrak{ad}_{\Ud(\vt)} (H) \Bigr> \\ 
    + \Bigl< \mathfrak{ad}_{\Vd(x)} \mathfrak{ad}_{\Ud(\vt)} (H) \Bigr>  - \braket{H} 
\end{multline}
where the sample index $i$ is ignored for simplicity, and the four middle auxiliary terms cancel each other. The purpose of the telescope expansion is to observe the penetration of data difference $\d x_j$ and parameter transition $\d \theta_j$ via the three pairs in (\ref{E: target loss 3}). For this, we utilize the Lie group adjoint action approximation given in Sec.~\ref{subsec: Adjoint approx}.  Knowing that (\ref{E:ad approx 2}) can be written as
\begin{equation}\label{E:ad approx 3}
 \mathfrak{ad}_{e^{i t \s}}(B) = B + it [\s, B] + \mathscr{O}(t^2)   
\end{equation}
(\ref{E: circuit connection}) gives us,
\begin{equation}\label{E:Ad V composition}
\begin{aligned}
    \mathfrak{ad}_{\Vd_j(\wtilde{x}_j)} &= \mathfrak{ad}_{\Vd_j(x_j)} \, \mathfrak{ad}_{\Vd_j(\d x_j)} = \mathfrak{ad}_{\Vd_j(\d x_j)} \, \mathfrak{ad}_{\Vd_j(x_j)}\\
    &=  \mathfrak{ad}_{\Vd_j(x_j)} + \frac{i}{2} \d x_j \, [\s_{k_j}, \mathfrak{ad}_{\Vd_j(x_j)}] + \mathscr{O}((\d x_j)^2)
    \end{aligned}
\end{equation}
where function compositions $\circ$ are automatically implied for the actions between the adjoint actions. 

With (\ref{E:Ad V composition}), the first pair of RHS (\ref{E: target loss 3}) yields
\begin{multline}\label{E: encoding circuits expansion 2}
\frac{i}{2} \, \d x_1 \, \Bigl< \left[ \s_{k_1}, \mathfrak{ad}_{\Vd_1(x_1)} \mathfrak{ad}_{\Vd_2(\wtilde{x}_2)} \cdots \mathfrak{ad}_{\Vd_d(\wtilde{x}_d)} \mathfrak{ad}_{\Ud(\wtilde{\vt})} (H) \right] \Bigr>  \\
+ \frac{i}{2} \, \d x_2 \, \Bigl< \mathfrak{ad}_{\Vd_1(x_1)} \circ \left[ \s_{k_2},  \mathfrak{ad}_{\Vd_2(x_2)} \cdots \mathfrak{ad}_{\Vd_d(\wtilde{x}_d)} \mathfrak{ad}_{\Ud(\wtilde{\vt})} (H) \right] \Bigr> \\
\vdots\\
+ \frac{i}{2} \, \d x_d \, \Bigl< \mathfrak{ad}_{\Vd_1(x_1)} \cdots \mathfrak{ad}_{\Vd_{d-1}(x_{d-1})} \circ \left[ \s_{k_d}, \mathfrak{ad}_{\Vd_d(x_d)} \mathfrak{ad}_{\Ud(\wtilde{\vt})} (H) \right] \Bigr> \\
+ \mathscr{O}((\d x)^2)
\end{multline}
Similarly, for variational circuits, we have
\begin{equation}\label{E:Ad U composition}
    \mathfrak{ad}_{\Ud_{\ell}(\wtilde{\vt}_{\ell})} = \mathfrak{ad}_{\Ud_{\ell}(\vt_{\ell})} + \frac{i}{2} \d \vt_{\ell} \, [\s_{k_{\ell}}, \mathfrak{ad}_{\Ud_j(\vt_{\ell})}] + \mathscr{O}((\d \vt_{\ell})^2)
\end{equation}
Using (\ref{E:Ad U composition}) for the second pair in RHS (\ref{E: target loss 3}) shows us how the VQC parameters react to the new domain change,
\begin{multline}\label{E: variational circuits expansion 2}
\frac{i}{2} \, \d \vt_1 \, \Bigl< \mathfrak{ad}_{\Vd(x)} \circ \left[ \s_{k_1}, \mathfrak{ad}_{\Ud_1(\vt_1)} \cdots \mathfrak{ad}_{\Ud_L(\wtilde{\vt}_L)} (H) \right] \Bigr>  \\
+ \frac{i}{2} \, \d \vt_2 \, \Bigl< \mathfrak{ad}_{\Vd(x)} \mathfrak{ad}_{\Ud_1(\vt_1)} \circ \left[ \s_{k_2}, \mathfrak{ad}_{\Ud_2(\vt_2)} \cdots \mathfrak{ad}_{\Ud_L(\wtilde{\vt}_L)} (H) \right] \Bigr> \\
\vdots\\
+ \frac{i}{2} \, \d \vt_L \, \Bigl< \mathfrak{ad}_{\Vd(x)} \mathfrak{ad}_{\Ud_1(\vt_1)} \ldots \mathfrak{ad}_{\Ud_{L-1}(\wtilde{\vt}_{L-1})} \circ \left[ \s_{k_L}, \mathfrak{ad}_{\Ud_L(\vt_L)} (H) \right] \Bigr> \\
+ \mathscr{O}((\d \vt)^2)
\end{multline}
Then it is noticed both  (\ref{E: encoding circuits expansion 2}) and (\ref{E: variational circuits expansion 2}) can be simplified as inner products $\langle r, \d x \rangle_{\mathbb{R}^d}$ and $\langle z, \d \vt \rangle_{\mathbb{R}^L}$, respectively, with two vectors $r:= (r_1, \ldots, r_d) \in \mathbb{R}^d$, $z := (z_1, \ldots, z_L) \in \mathbb{R}^L$ defined,
\begin{multline}\label{E: r component}
      r_j := \frac{i}{2} \Bigl< \mathfrak{ad}_{\Vd_1(x_1)} \cdots \mathfrak{ad}_{\Vd_{j-1}(x_{j-1})} \circ \\
      \left[ \s_{k_j}, \mathfrak{ad}_{\Vd_j(\wtilde{x}_j)} \cdots \mathfrak{ad}_{\Vd_d(\wtilde{x}_d)} \mathfrak{ad}_{\Ud(\wtilde{\vt})} (H) \right] \Bigr>
\end{multline}
\begin{multline}\label{E: z component}
  z_{\ell} := \frac{i}{2} \Bigl< \mathfrak{ad}_{\Vd(x)} \mathfrak{ad}_{\Ud_1(\vt_1)} \cdots \mathfrak{ad}_{\Ud_{\ell - 1}(\vt_{\ell - 1})} \circ \\
  \left[ \s_{k_{\ell}}, \mathfrak{ad}_{\Ud_{\ell}(\wtilde{\vt}_{\ell})} \cdots \mathfrak{ad}_{\Ud_{\ell}(\wtilde{\vt}_L)} (H) \right] \Bigr> 
\end{multline}
Note that it is the imaginary number $i$ in front of (\ref{E: r component}), (\ref{E: z component}) that makes $r_j, z_{\ell}$ all \emph{real}. The above circuit layer analysis converts the fine-tune loss (\ref{E: target loss}) into the problem of finding $\d \vt$ in a \emph{linear form},
\begin{equation}\label{E: finetune loss 3}
L(\wtilde{\vt}; \wtilde{\mcal{D}}) = \sum_{i=1}^N \Big\| \bigl< z, \d \vt \bigr>_{\mathbb{R}^L}(x^{(i)}) - q_i \Big\|^2
\end{equation}
with constant $q_i$ defined as,
\begin{equation}\label{E: VQC residue}
    q_i = \d y^{(i)} - \bigl< r, \d x^{(i)} \bigr>_{\mathbb{R}^d} + y^{(i)} - \braket{H} (x^{(i)}; \vt)
\end{equation}
This quantity is called the \emph{VQC transfer residue} that bears a certain meaning in understanding the VQC transfer mechanism.

\section{Interpretations}\label{sec: interpretations}

As (\ref{E: finetune loss 3}) and (\ref{E: VQC residue}) depict how a VQC adjusts to the slight domain mismatch, an \emph{optimal} finetune can be analytically derived via the Moore-Penrose pseudo inverse with $q :=(q_1, \ldots, q_N)$ defined,
\begin{equation}\label{E: QVA}
    \d \vt^* =  (z^T \cdot z)^{-1} \, z^T \cdot q
\end{equation}
This provides a one-shot parameter shift for VQC to adapt to $\wtilde{\mathcal{D}}$ promptly, see Fig.~\ref{fig: Unitary param transition}. Subsequently, (\ref{E: QVA}) reveals that
\begin{equation}\label{E: q meaning}
    \begin{aligned}
    &\| q \| \to 0, \quad \d \vt^* \to 0 \quad \text{(nothing to adapt)}\\
    &\| q \| \gg 1, \quad \d \vt^* \gg 1 \quad \text{(much to adapt)}
\end{aligned}
\end{equation}
to justify the name \emph{VQC transferal residue} for quantity $q$.

\subsection{VQC Transferal residue}
We look further into the composition of $q$, which consists of two pairs,
\[
    q_i = \underbracket{ \d y^{(i)} - \bigl< r, \d x^{(i)} \bigr>_{\mathbb{R}^d} }_{\text{domain mismatch}(\Delta)} + \overbracket{y^{(i)} - \braket{H} (x^{(i)}; \vt)}_{\text{pretrain error} (\star)}  \tag{\ref{E: VQC residue}}
\]
where the first pair $(\Delta)$ describes the \emph{mismatch} of two domains, and the other pair evaluates the error from the pretrained VQC model. Indeed, (1) a perfect pretrain $f$ with $f( x_i ) \equiv y_i$ gives $(\star) = 0$, and (2) for two domains perfectly match $\mcal{D} = \mcal{\wtilde{D}}$, we have $\delta x_i = \delta y_i = 0$ and $(\Delta)=0$.

Under such perfect circumstances, $q_i \equiv 0$ and there is \emph{nothing new} for $f$ to learn or adapt in $\mcal{\wtilde{D}}$ by (\ref{E: QVA}). On the other hand, for an ill-pretrained $f$ with large domain difference $\| \delta x_i \| $, $\| \delta y_i \| $,  $q_i$ becomes large too. This indicates $\d \vt^*$ has \emph{much to adapt} in $\mcal{\wtilde{D}}$. Therefore, the transferal residue characterizes the amount of new knowledge needed to be learned in the new domain, and hence the name.

\subsection{Knowledge transfer in VQC} 
As $q$ reflects the amount of new knowledge left to be digested in $\mcal{\wtilde{D}}$, the term $ \bigl< r, \d x^{(i)} \bigr>$ is observed to automatically rise to reduce new label mismatch $\d y$, which can be regarded as a \emph{self-correction (self-adaptation)} to the new domain. Indeed, the self-correction vanishes when $\mcal{D} = \mcal{\wtilde{D}}$ or $\delta x^{(i)} =0$. 

As the pretrained VQC $f$ was trained with the old data, the term $r$ in (\ref{E: r component}) carries \textit{previous domain knowledge} via old circuits configurations $U_1(\vt_1), \ldots U_{L}(\vt_L)$ to resolve new data mismatch. This formulation clearly indicates how the domain transition is performed to provide an intuition for the VQC transfer learning. Similar interpretations on network-based models can be found in \cite{tseng2023interpretations}. However, the form of (\ref{E: r component}) and (\ref{E: z component}) is particular to the VQC structure. In fact, the above analysis includes four types of domain transitions,

\textbf{Type 1 ($\delta x_i = \delta y_i = 0$):} This implies $\mcal{D} = \mcal{\wtilde{D}}$ requiring no adaptations, as $q_i= 0 $ leads to $\d \vt = 0 $  by (\ref{E: QVA}).

\textbf{Type 2 ($\delta x_i \neq 0$, $\delta y_i = 0$):} This corresponds to the regular data augmentation, as there are new inputs $\wtilde{x}_i \neq x_i$ giving the same label $y_i$.

\textbf{Type 3 ($\delta x_i = 0$, $\delta y_i \neq 0$):} This case is to augment mislabelled data since the new input $\wtilde{x}_i$ is the same as $x_i$, yet labels differ $\wtilde{y_i} \neq y_i$.

\textbf{Type 4 ($\delta x_i \neq 0$, $\delta y_i \neq 0$):} This is a general case that contains all the types above.

\section{Experiment}
We validate the optimal adaptation solution in Sec.~\ref{subsec: VQC fine-tune error} with 1-qubit demonstration.

\begin{figure}[htb]
\vskip -0.05in
\begin{center}
\includegraphics[width=1\columnwidth]{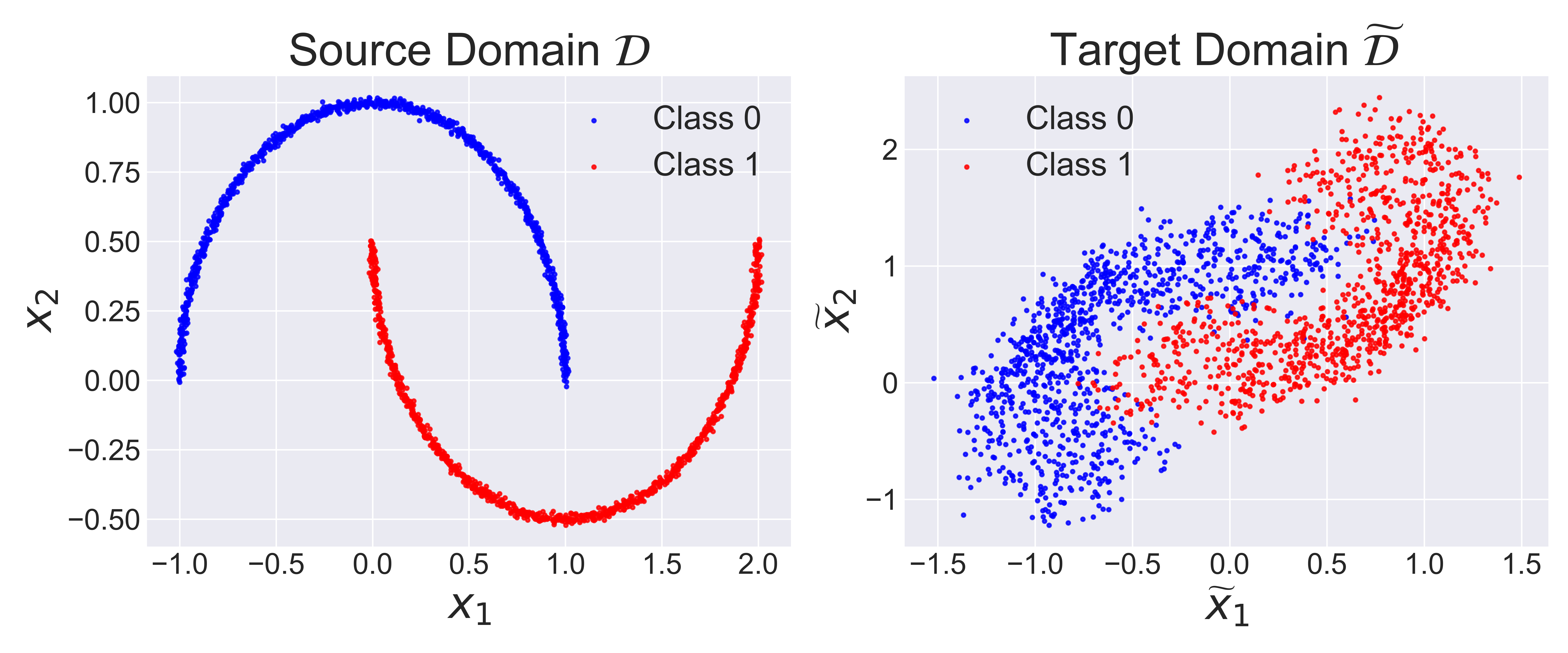}
\vspace*{-7mm}
\caption{Transfer learning of VQC from $\mathcal{D} \to \wtilde{\mathcal{D}}$.}
\label{fig: datasets}
\end{center}
\vskip -0.2in
\end{figure}

A pair of ``\emph{two moons}" datasets (with 2000 samples each) are used to conduct transfer learning, one served as the \emph{source domain} $\mathcal{D}$; the other as the \emph{target domain} $\wtilde{\mathcal{D}}$, see Fig.~\ref{fig: datasets}. A pretrained model $f$ is well-trained (attaining accuracy $81.5\%$) on $\mathcal{D}$. However, the accuracy of $f$ drops promptly to $49.8\%$ on $\wtilde{\mathcal{D}}$ such that the domain adaptation is necessitated.

\begin{figure}[htb]
\vskip -0.0in
\begin{center}
\includegraphics[width=1\columnwidth]{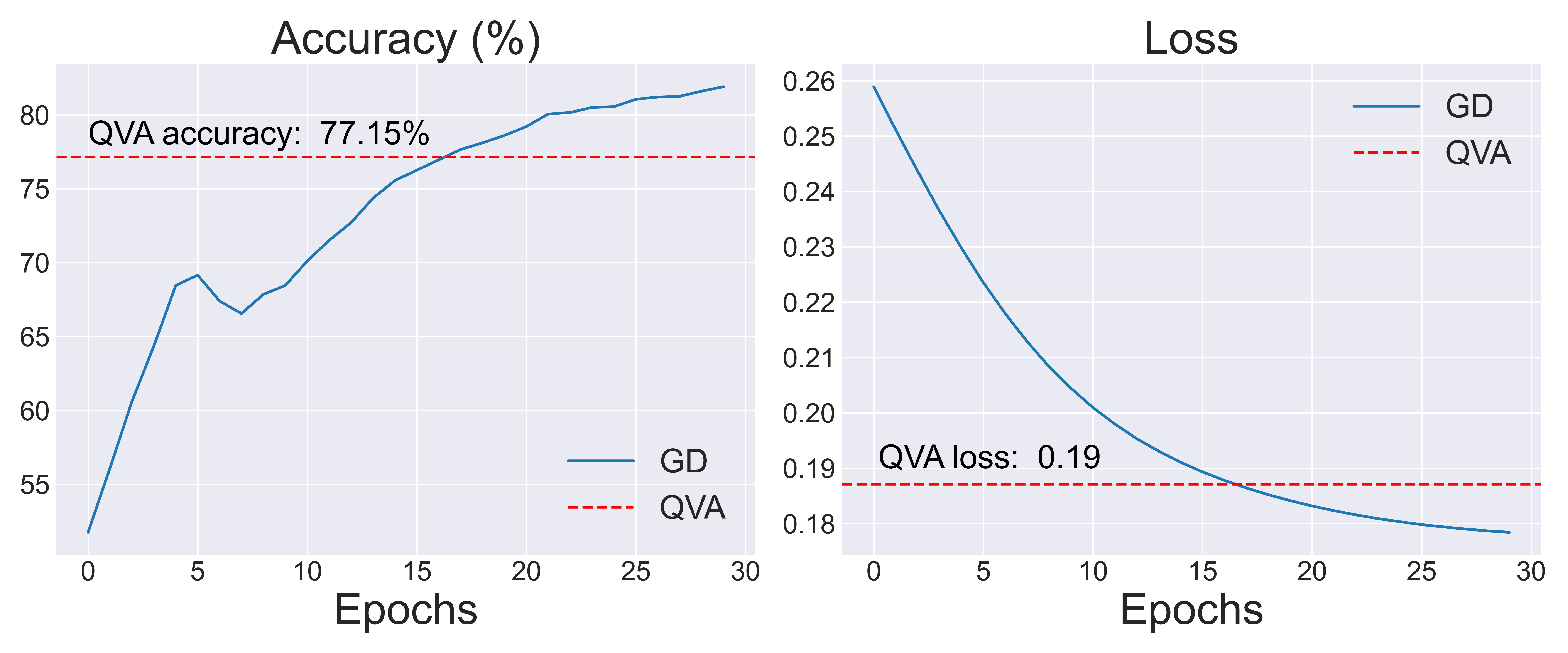}
\vspace*{-6mm}
\caption{Training curve and accuracy during adaptation.}
\label{fig: results}
\end{center}
\vskip -0.1in
\end{figure}

Consequently, our one-shot fine-tune solution (\ref{E: r component})$\sim$(\ref{E: QVA}) (named \emph{Quantum Variational Analysis}, \textbf{QVA} for short) is conducted to compare with the conventional Gradient Descent (GD) method. Fig.~\ref{fig: results} shows the GD training loss and accuracy over 30 epochs. As QVA requires no training, it is represented by one horizontal line. The results indicate that the QVA directly yields $77.2\%$ accuracy on $\wtilde{\mathcal{D}}$ without iterative GD process. While GD continues to improve with more training iterations, our approach provides prompt adaptations competing with the first 17 epochs, offering an efficient alternative.

\section{Conclusion}
This study investigates transfer learning in VQC, using algebraic estimations to unveil the underlying mechanism. Our analytical computation yields a direct optimal fine-tune solution under similar domain conditions without the need for iterative gradient descent. The theoretical description additionally provides an intuitive interpretation of how VQC attempts to resolve the domain mismatch and adaptation to new data. This framework also reveals the unique characteristics of transfer learning mechanisms pertaining to VQCs, offering insights into quantum models.

\clearpage
\bibliographystyle{IEEEtran}
\bibliography{references,bib/qc,bib/vqc,bib/qml_examples,bib/transfer_learning,bib/qt}

\end{document}